\documentclass[11pt]{cernrep}
\usepackage{graphicx}
\input{epsf} 
\usepackage{here}
\usepackage{cite}

\begin{document}
\bibliographystyle{unsrt}
\title{ A COMPARISON OF PREDICTIONS FOR SM HIGGS BOSON PRODUCTION AT THE LHC}
\author{C. Balazs$^1$, M. Grazzini$^2$, J. Huston$^3$, A. Kulesza$^4$, I. Puljak$^5$}
\institute{$^1$HEP Division, Argonne National Laboratory, 9700 Cass Ave., Argonne IL 60439
USA\footnote{ ANL is supported by the U.S. Department of Energy HEP Division
under the contract W-31-109-ENG-38.}\\
$^2$Department of Physics, CERN, Theory Division, CH--1211 Geneva 23, Switzerland\\
$^3$Department of Physics and Astronomy, Michigan State 
University, USA\\ 
$^4$Institut fur Theoretische Teilchenphysik, Universitaet Karlsruhe, Germany\\
$^5$FESB, University of Split, Split, Croatia}
\maketitle

\def\ltap{\raisebox{-.4ex}{\rlap{$\,\sim\,$}} \raisebox{.4ex}{$\,<\,$}} 
\def\gtap{\raisebox{-.4ex}{\rlap{$\,\sim\,$}} \raisebox{.4ex}{$\,>\,$}} 












\section{INTRODUCTION}

The dominant mechanism for the production of a SM
Higgs boson at the
LHC is gluon-gluon fusion through a heavy (top) quark loop.
For this reason this channel
has attracted a large amount of theoretical attention~\cite{Cavalli:2002vs}.
Recently, the total cross section has been calculated to NNLO 
in the strong coupling constant $\alpha_s$
(i.e. at order ${\cal O}(\alpha_s^4)$)
\cite{Catani:2001ic,Harlander:2001is,Harlander:2002wh,Anastasiou:2002yz,Ravindran:2003um}
and also contributions from multiple soft gluon emission have been
consistently included to NNLL accuracy \cite{Catani:2003zt}.
In addition to the size of the total
rate, a knowledge of the shape of the Higgs boson $p_T$ distribution
is essential for any search and analysis strategies at the LHC. In
particular, the $p_T$ distribution for the Higgs boson is expected
to be harder than the one of its corresponding backgrounds.
The Higgs boson $p_T$ distribution has been computed with LL parton shower
Monte Carlos (HERWIG~\cite{Corcella:2002jc}
and PYTHIA~\cite{Sjostrand:2003wg}),
and through
various resummed calculations. The latter techniques are the more powerful
ones, but it is primarily the former that experimentalists at the LHC
have to rely upon, because of their flexibility in allowing to test
the effects of the various kinematic cuts which may optimize search 
strategies.

In the kinematic region $p_T^2\ll m_H^2$, where most of the events
are expected,
large logarithmic corrections
appear of the form $\alpha_s^n \ln^{m}m_H^2/p_T^2$ that spoil the validity
of the fixed order perturbative expansion.
The $p_T$ distribution can be written as
\begin{eqnarray}
\label{deco}
\frac{d\sigma}{dp_T^2} = \frac{d\sigma^{\rm res.}}{dp_T^2} +
\frac{d\sigma^{\rm fin.}}{dp_T^2}\, .
\end{eqnarray}
The first term contains all
logarithmically-enhanced contributions and requires their resummation to all
orders. The second term is free from logarithmically-enhanced contributions
and can be evaluated at fixed order in perturbation theory.
The method to perform the all-order resummation is well known:
to correctly take into account momentum conservation
the resummation must be performed in the
impact parameter ($b$) space \cite{Parisi:1979se,Dokshitzer:1980hw}.
The large logarithmic contributions are exponentiated in the
Sudakov form factor, which
in the CSS \cite{Collins:1985kg}
approach takes the form
\begin{eqnarray}
S_c =
\int_{b_0^2/b^2}^{m_H^2} \frac{d \mu^2}{\mu^2}
\left[
A_c \left( \alpha_s(\mu) \right) \ln \left( \frac{m_H^2}{\mu^2} \right) +
B_c \left( \alpha_s(\mu) \right)
\right] ,
\label{Eq:Sudakov}
\end{eqnarray}
where $b_0=2e^{-\gamma}$ and $c=q,g$.
The $A_c$ and $B_c$ functions
are free of large logarithmic corrections
and can be computed as expansions in the strong
coupling constant $\alpha_s$:
\begin{eqnarray}
A_c(\alpha_s) = &&
\sum_{n=1}^\infty
\left( \frac{\alpha_s}\pi \right)^n A_c^{(n)}, \\
B_c(\alpha_s) = &&
\sum_{n=1}^\infty
\left( \frac{\alpha_s}\pi \right)^n B_c^{(n)}.
\end{eqnarray}
The functions $A_c$ and $B_c$ control soft and flavour-conserving collinear radiation
at scales $1/b \ltap \mu \ltap m_H$.
Purely soft radiation at a very low scales $\mu \ltap 1/b$ cancels out because
the cross section is infrared safe and
only purely collinear radiation
up a scale $\mu \sim 1/b$ remains, which is taken into account by
the coefficients
\begin{equation}
C_{ab}(\alpha_s,z) =
\sum_{n=1}^\infty
\left( \frac{\alpha_S}\pi \right)^n C_{ab}^{(n)}(\alpha_s,z).
\end{equation}
Beyond NLL accuracy, to preserve the process independence of
the resummation formula,
an additional (process dependent)
coefficient $H$ is needed \cite{Catani:2000vq},
which accounts for
hard virtual corrections and has an expansion
\begin{equation}
H_c(\alpha_s)=1+\sum_{n=1}^{\infty}
\left( \frac{\alpha_s}\pi \right)^n H_c^{(n)}.
\end{equation}
In the case of Higgs boson production through $gg$ fusion, the relevant
coefficients $A_g^{(1)}$, $A_g^{(2)}$ and $B_g^{(1)}$
are known \cite{Catani:1988vd}
and control the resummation up to NLL accuracy
\footnote{There are two different classification schemes of the LL, NLL, NNLL, etc
terms and their corresponding B contents. Here we use the most popular
scheme. Another is discussed in Ref.~\cite{Giele:2002hx}.}.
The NNLL coefficients $C^{(1)}_{ab}$ and $H_g^{(1)}$ are also known
\cite{Kauffman:1992cx,Catani:2000vq}.
The NNLL coefficient $B_g^{(2)}$
has been computed in Refs.~\cite{deFlorian:2000pr,deFlorian:2001zd},
whereas $A_g^{(3)}$ is not yet known exactly.
In the following we assume that its value
is the same that appears in threshold resummation \cite{Vogt:2000ci}.

\section{PREDICTIONS FOR $p_T$ SPECTRA AND COMPARISONS}

In the 1999 Les Houches workshop,
a comparison~\cite{Catani:2000zg,Balazs:2000sz} of the
HERWIG and PYTHIA (2 versions) predictions for the Higgs boson $p_T$
distribution with those of a $p_T$ resummation program
(ResBos~\cite{Balazs:1997xd,Balazs:2000wv}) was carried out. This comparison was
continued in the 2001 workshop and examined the impact
of the $B^{(2)}$ coefficient~
\cite{Cavalli:2002vs}.
In the meantime, a
number of new theoretical predictions have become available, both from
resummation and from the interface of NLO calculations 
with parton shower Monte Carlos. For these proceedings,
we have carried out a comparison of most of the
available predictions for the Higgs boson $p_T$ distribution at the LHC. We
have used a Higgs boson mass of 125 GeV
and either  the MRST2001 or the CTEQ5M pdf's.
The difference between the two pdf's for the production of a 125 GeV
mass Higgs boson is  of the order of a few percent.
Before comparing the different predictions, we comment on the various approaches in turn.

Parton shower MC programs such as HERWIG, which implements
angular ordering exactly, implicitly
include the $A^{(1)}$, $A^{(2)}$ and $B^{(1)}$ coefficients
and thus correctly sum the LL
and part of the N$^k$LL contributions. However, in the most straightforward
implementations, MC cannot correctly treat hard radiation. 
By contrast, the PYTHIA MC,
which does not provide an exact implementation
of angular ordering, has
a hard matrix element correction
\footnote{Very recently hard matrix element corrections for Higgs productions
have been implemented in HERWIG as well \cite{Corcella:2004fr}.}.
Recently, an approach to match NLO calculations to
parton showers generators, MC@NLO \cite{Frixione:2002ik,Frixione:2003ei},
has been proposed, and applied, amongst the other, to Higgs production.
This method joins the virtues of
NLO parton level generators
(correct treatment of hard radiation, exact NLO normalization)
to the ones of MC. It thus can be compared
to a resummed calculation at NLL+NLO accuracy.

As far as resummed calculations are concerned,
we first consider two implementation of the CSS approach.
The ResBos code
includes the $A_c^{(1,2,3)}$, $B_c^{(1,2)}$ and $C_{ab}^{(1)}$
coefficients
in the low-$p_T$ region and matches this to the NLO distribution at high $p_T$.
NNLO effects at high $p_T$ are approximately taken into account by
scaling the second term in Eq.~(\ref{deco}) with a K-factor.
The matching is performed through a switching procedure whose uncertainty
will be considered in the following.
The calculation of Berger and Qiu~\cite{Berger:2002ut} also performs a $p_T$
resummation in $b$ space and is accurate to NLL. The coefficient
$B^{(2)}$ is included but the matching is still to NLO.
Note that in both these approaches the integral of the spectrum
is affected by higher-order contributions
included in a non-systematic manner
whose effect is not negligible for Higgs production.

The prediction by Bozzi, Catani, de Florian and Grazzini~\cite{Bozzi:2003jy} (labeled Grazzini et al. in the following)
is based on an implementation of the $b$-space
formalism described in \cite{Catani:2000vq,Bozzi:2003jy}. The calculation has
the highest nominal accuracy since it matches NNLL resummation at small
$p_T$ to the NNLO result at high $p_T$\cite{deFlorian:1999zd}.
This approach includes the coefficients
$C_{ab}^{(2)}$ and $H_g^{(2)}$ in approximated form.
The main differences with respect to the standard CSS approach
are the following.
A unitarity constraint is imposed, such that the total cross section at the
nominal (NNLO) accuracy is exactly recovered upon integration.
A study of uncertainties from missing higher order contributions
can be performed as it is normally done in fixed order calculations, that is,
by varying renormalization and factorization scales around the central value,
that is chosen to be $m_H$.

Finally, we discuss the $p_T$ distribution of
Ref.~\cite{Kulesza:2003wn} (Kulesza et al.).
This is obtained using a
joint resummation formalism, by which
both threshold and low-$p_T$
logarithmic contributions are resummed to all orders. This approach
has been formally developed to NLL accuracy, but the NNLL coefficients
$A^{(3)}, B^{(2)}, C^{(1)}$ and $H^{(1)}$ can also be incorporated.
The matching is still performed to NLO.
Even though a low mass Higgs boson
at the LHC is produced with relatively low $x$ partons, threshold effects
can still be significant due to the large color charge in the
$gg$ initial state as well as steep $x$ dependence of the gluon
distribution functions at low $x$. This leads to an increased sensitivity
to Sudakov logarithms associated with partonic threshold for gluon-induced
processes, as shown in Ref.~\cite{Catani:2003zt}. 

It is known that the low-$p_T$ region is sensitive to
non-perturbative effects. These are expected to be less
important in the gluon channel due to the larger colour charge
of the $gg$ initial state \cite{Balazs:2000sz}.
Different treatments of non-perturbative effects are included in
the ResBos, Berger et al. and Kulesza et al calculation,
whereas Grazzini et al. prediction is purely perturbative.

\begin{figure}[htb]
\begin{center}
\begin{tabular}{c}
\epsfxsize=12truecm
\epsffile{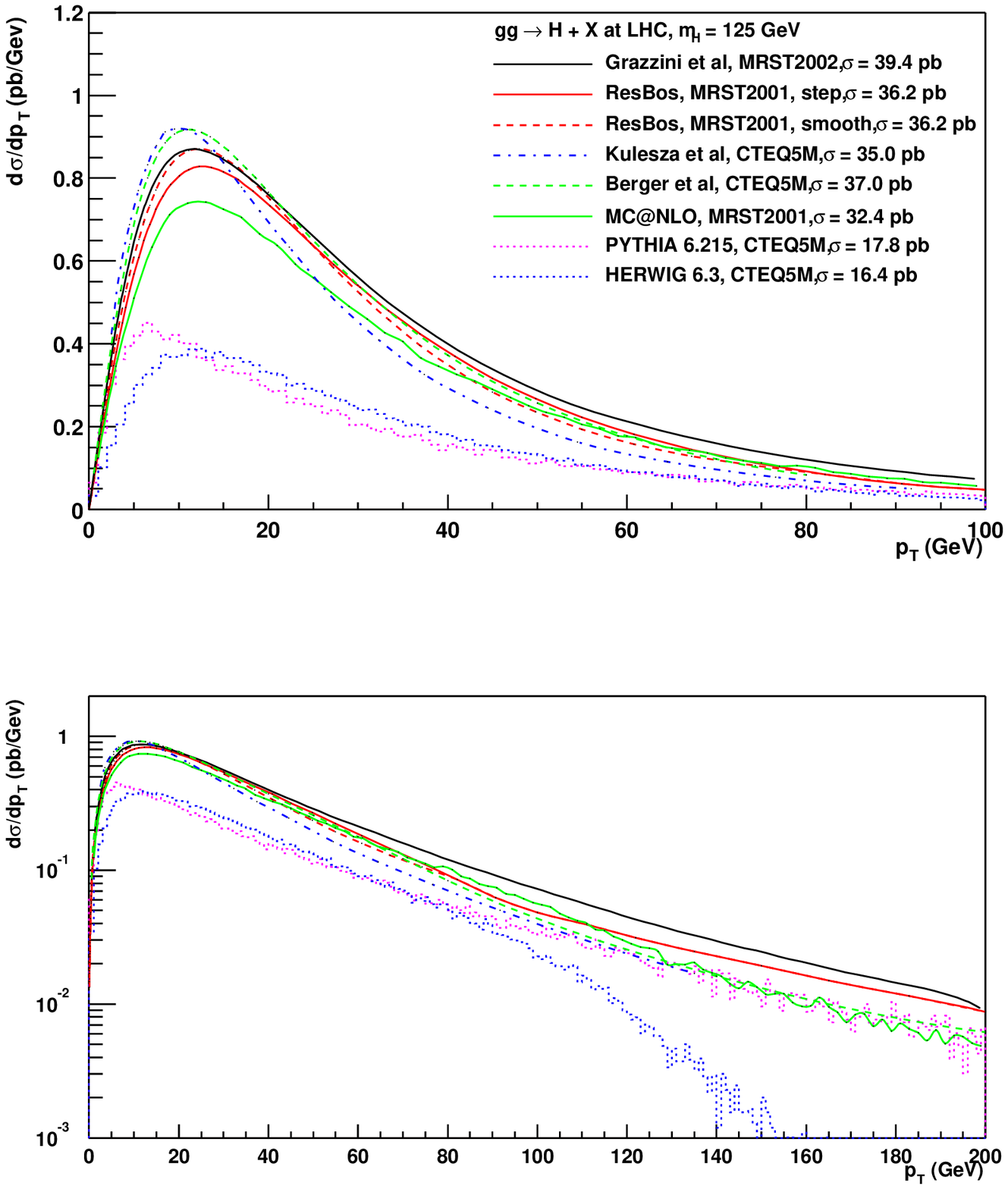}\\
\end{tabular}
\end{center}
\vspace*{-5mm}
\caption{The absolute predictions for the production of a 125 GeV mass  Higgs boson at the LHC.
\label{fig:higgs_abs_norm}}
\end{figure}

The absolute predictions for the cross sections
are shown in Figure~\ref{fig:higgs_abs_norm}.
All curves are obtained in the $m_{\rm top}\to\infty$ limit.
HERWIG and PYTHIA cross sections are significantly smaller than
the other predictions, their normalization being fixed to LO.
In the high-$p_T$ region, the HERWIG prediction drops quickly due to the
lack of hard matrix element corrections. PYTHIA, in
contrast, features the hard matrix element corrections.
We also note that PYTHIA prediction is significantly
softer than all the other curves, and thus
its overall shape is fairly different
from all the other predictions.

The MC@NLO cross section, about 32.4 pb, is roughly
twice that of the HERWIG and PYTHIA predictions, being fixed to the NLO total
cross section.

Two predictions ({\em step, smooth}) are shown for ResBos 
which differ in the manner in which the matching at high $p_T$
is performed. Their difference can be considered as
an estimate of the ambiguity in the switching procedure.
The two curves correspond to the same total cross section of about 36.2 pb,
which is about 8 \% higher than the NLO cross section.
This is the effect of the higher-order terms
that enter the prediction for the total rate in
the context of the CSS approach.
A slightly softer curve is obtained by Berger and Qiu.
The predicted cross section (37 pb) is close to that of ResBos. 

The Grazzini et al. prediction has an integral of about 39.4 pb, which
corresponds to the total cross section at NNLO. Contrary to what is done
in Ref.~\cite{Bozzi:2003jy}, here the curve is obtained with
MRST2002 NNLO partons and three-loop $\alpha_s$.
The difference with the result obtained with MRST2001 NNLO PDFs
is completely negligible.

Concerning the Kulesza et al curve,
the subleading terms associated with low $x$ emission
(i.e. in the limit opposite to partonic threshold) and of which only a
subset is included in the joint resummation formalism, play an important role
numerically. As a result, the total cross section turns out to be 35 pb,
about $10\%$ lower than the pure threshold result, which is 39.4 pb
 \cite{Kulesza:2003wn}.

\begin{figure}[htb]
\begin{center}
\begin{tabular}{c}
\epsfxsize=12truecm
\epsffile{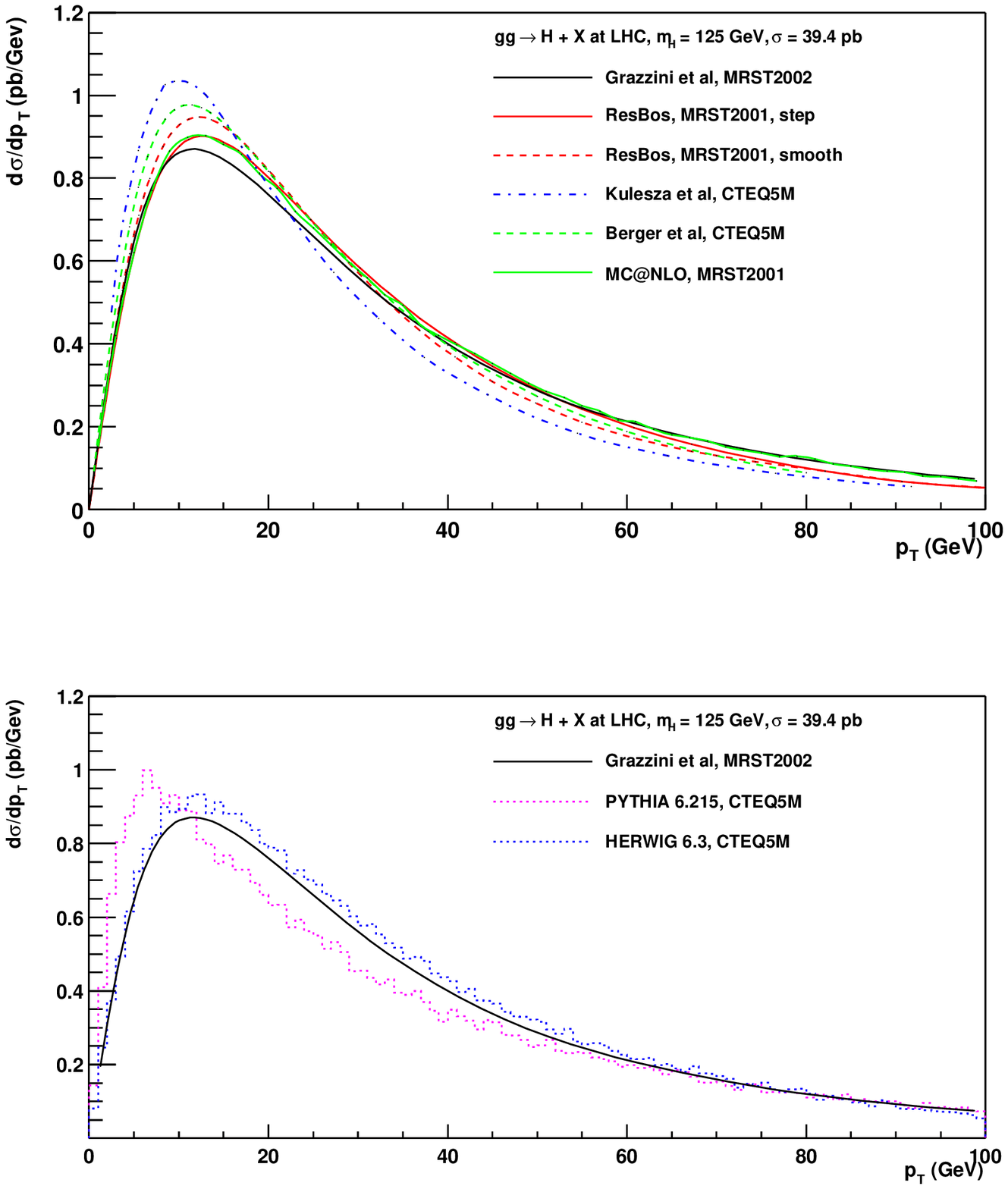}\\
\end{tabular}
\end{center}
\vspace*{-5mm}
\caption{\label{fig:higgs_no_norm}The predictions for the production of a 125 GeV mass  Higgs boson at the LHC, all normalized to the same cross section for better shape comparison.}
\end{figure}

We now want to examine in more detail the relative shapes of the
predictions plotted in Figure.~\ref{fig:higgs_abs_norm}.
In Figure.~\ref{fig:higgs_no_norm} all the predictions are normalized
to the Grazzini et al. cross section of 39.4 pb.

In the region of small and moderate $p_T$ (say, smaller than 100 GeV) 
all of the predictions are basically consistent with each other, with
the notable exception of PYTHIA, which predicts a much softer spectrum.
The curve of Kulesza et al. is also softer than the others.

For larger $p_T$, HERWIG gives unreliable predictions, since the
transverse momentum is generated solely by means of the parton shower,
and therefore it lacks hard matrix element effects.
The Grazzini et al. and ResBos curves are harder than
MC@NLO for large $p_T$. There are two reasons for this. Grazzini 
et al. implement  the NNLO matrix elements exactly, corresponding to the
emission of two real partons accompanying
the Higgs in the final state \cite{deFlorian:1999zd};
ResBos mimics these contributions, by multiplying the NLO matrix elements
by the K factor. MC@NLO, on the other hand, contains only NLO matrix
elements (one real parton in the final state). Secondly, Grazzini et al. 
and ResBos choose the renormalization and factorization scales equal to $m_H$,
whereas in MC@NLO these scales are set equal to the transverse mass of the
Higgs, $\sqrt{m_H^2+p_T^2}$. The difference is small at the level of
total rates, but it is not negligible in the tail of the $p_T$ distribution.

\section{CONCLUSIONS}

Up to now, the ATLAS  and CMS  experiments have relied
primarily on the predictions from HERWIG and PYTHIA in designing both
their experiments as well as defining their search and analysis
strategies. In the last few years, a number of tools for and predictions 
of the Higgs boson cross section at the LHC have become available, with 
the inclusion of beyond-the-leading-order effects at different level of 
accuracy. In the case of total rates, NNLO results have recently become
available; their consistent inclusion in experimental analysis will
allow to further decrease the estimated lower bound on the integrated 
luminosity to be collected for discovery.

In this contribution, we primarily 
focused on the predictions for the $p_T$ spectrum, comparing the results
of Monte Carlos with those obtained with analytically-resummed calculations.
In contrast to the situation in 1999, all of the predictions, with the
exception of PYTHIA, result in the same general features, most notably in
the position of the peak. However, differences do arise, because of 
different treatments of the higher orders. It is an interesting question
beyond the scope of this review that of whether these differences are
resolvable at the experimental level, which may lead to modify the strategy
for searches. In order to answer this, studies including realistic 
experimental cuts must be performed with the newly available tools.

\bibliography{higgscomparison}
\end{document}